# PERFORMANCE ANALYSIS OF MIMO RADAR WAVEFORM USING ACCELERATED PARTICLE SWARM OPTIMIZATION ALGORITHM

B. Roja Reddy[1] and Uttara Kumari .M[2]

[1]Department of Telecommunication & Electronics and Communication Engineering,
rojareddyb@gmail.com
[2] RVCE, Mysore Road Bangalore, Karnataka, India.
uttarakumari@rvce.edu.in

## ABSTRACT

*The Accelerated Particle Swarm Optimization Algorithm is promoted to numerically design orthogonal Discrete Frequency Waveforms and Modified Discrete Frequency Waveforms (DFCWs) with good correlation properties for MIMO radar. We employ Accelerated Particle Swarm Optimization algorithm (ACC_PSO), Particles of a swarm communicate good positions, velocity and accelerations to each other as well as dynamically adjust their own position, velocity and acceleration derived from the best of all particles. The simulation results show that the proposed algorithm is effective for the design of DFCWs signal used in MIMO radar.*

## 1. INTRODUCTION

Recently, the concept of MIMO radars has drawn considerable attention due to their ability to image a target from a variety of angles, thus improving the information received from the target and capability of modern radars in terms of detection, identification and classification of target under surveillance [1]. Multiple input and multiple output system can transmit multiple linearly independent probing signals via its antennas MIMO systems have been deemed as efficient spatial multiplexers and ultimately a suitable strategy to ensure high-rate communications on wireless channels [2].

One of the important properties for MIMO radar is the resolution performance enhancement. The range resolution can be significantly improved by using very short pulses. This results in the decrease in received signal to noise ratio. To increase signal to noise ratio various Pulse compression techniques were developed. Although there are analog pulse compressions techniques, digital pulse compression is more popularly used. These include frequency codes, binary phase codes, and poly phase codes. The performance criteria of a pulse compression sequence are judged by their autocorrelation properties [3].

The orthogonally coded waveforms are transmitted from each antenna because of their low probability to intercept, their ability to offer potential performance improvement in complex propagation conditions and being uncorrelated. The orthogonal waveforms used by the MIMO radar systems must be carefully designed to avoid the self-interference and detection confusion. For high range resolution and multiple target resolution, the aperiodic autocorrelation functions of sequences should have low of peak sidelobes level. Design of Orthogonal code sets with low Autocorrelation side lobe peaks (ASP) and cross correlation peaks (CP) is crucial for the implementing MIMO radar systems.

There are several types of waveforms to design orthogonal waveforms, such as polyphase waveform Discrete Frequency-coding Waveform (DFCW) which has large compressed ratio. The polyphase compression ratio is relatively small when compared with DFCW. The Autocorrelation Sidelobes Peak (ASP) level of discrete frequency-coding waveform with fixed frequency pulses is very large. By replacing fixed frequency pulse with linear Frequency Modulation Pulses can lower ASP but the grating lobes will appear in the range response if T$\Delta$f >1 [3]. The relationship between subpulse duration T, frequency step $\Delta$f and LFM bandwidth B is set to make the grating lodes disappear [3].

Deng [4] has proposed simulated annealing algorithm to design of DFCW and got good results. Liu has proposed an approach using optimization Genetic Algorithm (GA) [3] technique and a Modified Genetic Algorithm (MGA) [5] technique to design multiple orthogonal discrete frequency-coding sequences with good aperiodic correlation.

In this paper, we employ ACC_PSO to design discrete frequency-coding waveform to obtain good correlation properties. To achieve this object the cost function was designed based on Peak Side lobe level Ratio and Integrated Side lobe Level Ratio. This algorithm as an optimization engine to obtain an optimal solution to this problem and also stabilizes to the solution in considerably lesser computational efforts. In this algorithm the Particles of a swarm communicate good positions to each other as well as dynamically adjust their own position, velocity and acceleration derived from the best position of all particles.

The rest of paper is organized as follows. In section 2 we formulate the waveform design using DFCW. In section 3 we introduce ACC_PSO to numerically optimize DFCW. Design results from proposed algorithm in section 4. Finally some conclusions are drawn in section 5.

## 2. THE WAVEFORM DESIGN

Linear Frequency Modulation (LFM) the most popular pulse compression method. The basic idea is to sweep the frequency band B linearly during the pulse duration T. B is the total frequency deviation and the time bandwidth product of the signal is BT. The spectral efficiency of the LFM improves as the time-bandwidth product increases, because the spectral density approaches a rectangular shape. Here we consider the sequence length of each waveform (N) as 128 and Number of antennas (L) as 3.

### 2.1. DFCW with Frequency Hopping

The DFCW_FH waveform is defined as [4]

$$S_p(t) = \sum_{n=0}^{N-1} A(t) e^{j2\pi f_n^p t} \tag{1}$$

Where $A(t) = \begin{cases} 1/T \longrightarrow 0 \le t \le T \\ 0 \longrightarrow elsewhere \end{cases}$

and p= 1, 2, …, L T is the subpulse time duration. N is the number of subpulse that are continuous with the coefficient sequence $\{n_1, n_2, n_3, \ldots\ldots\ldots n_N\}$ with unique permutation of sequence $\{0.1, 2, 3, \ldots\ldots\ldots N-1\}$. $f_n^p$ = n .$\Delta$f is the coding frequency of subpulse n of waveform p in the waveform DFCW_FH. $\Delta$f is the frequency step.

## 2.1. DFCW with Linear Frequency Modulation

By adding LFM to the DFCW_FH, the DFCW-LFM waveform is defined as [2]

$$S_p(t) = \sum_{n=0}^{N-1} e^{j2\pi f_n^p(t-nT)} . e^{j\pi kt^2}$$

Where $A(t) = \begin{cases} 1/T \longrightarrow 0 \le t \le T \\ 0 \longrightarrow elsewhere \end{cases}$

$$(2)$$

Where k is the frequency slope, $k = \dfrac{B}{T}$ .

## 2.3. Modified DFCW with Linear Frequency Modulation

The modified DFCW waveform is proposed in this paper and is defined as

$$S_p(t) = \sum_{n=0}^{N-1} e^{j2\pi f_n^p(t-nT)} . e^{j\pi kt^3}$$

Where $A(t) = \begin{cases} 1/T \longrightarrow 0 \le t \le T \\ 0 \longrightarrow elsewhere \end{cases}$          (3)

The relationship between subpulse duration T, frequency step Δf and LFM bandwidth B is set to make the grating lobes disappear, it also lower the ASP.The B and T for each pulse remains a constant. The Δf values are called frequency steps. Each LFM pulse has a different starting frequency. The choice of BT, T.Δf and B/Δf values are crucial for the waveform design. Different lengths of firing sequence, N, have different values for each of the above mentioned parameters [3].

Table1.   Length of firing sequences & related parameters

| N | B.T | B/Δf | T.Δf |
|---|-----|------|------|
| 8 | 18 | 6 | 3 |
| 16 | 36 | 12 | 3 |
| 32 | 72 | 24 | 3 |
| 64 | 144 | 48 | 3 |
| 128 | 288 | 96 | 3 |

The performance analysis is expressed in terms of Peak Side Lobe Ratio and integrated Sidelobe Level Ratio. The PSLR is a ratio of the peak sidelobe amplitude to the main lobe peak amplitude and is expressed in decibels. The autocorrelation and Cross correlation PSLR are defined as [8].

$$PSLR_A = 20 \log_{10} \left\{ \frac{\max_n \in sidelobe \ |A(S_1, n)|}{\max_n \in mainlobe \ |A(S_1, n)|} \right\}$$

$$PSLR_C = 20 \log_{10} \left\{ \frac{\max_n \in sidelobe \ |C(S_p, S_q, n)|}{\max_n \in mainlobe \ |C(S_p, S_q, n)|} \right\} where q \ne p \qquad (4)$$

The ISLR is a ratio of the integrated energy of all side lobes which spread across the whole time domain to the integrated energy of the main lobe in the pulse compression function.

$$ISLR_S = 20 \log_{10} \left\{ \frac{\sum onlySidelobe \ |A(S_p, n)| \sum |C(S_p, S_q, n)|}{\sum only mainlobe \ |A(S_p, n)|} \right\} where q \ne p \qquad (5)$$

The cost function can be written as

$$CF = \sum_{l=1}^{L} \left( \min\left(PSLR_{Al}\right) + \min\left(PSLR_{Cl}\right)\right) + \lambda \cdot \sum_{l=1}^{L} ISLR_{sl} \qquad (6)$$

where $\lambda$ is the relative weigh assigned to the ISLR and PSLR.

## 3. ACCELERATED PARTICLE SWARM OPTIMIZATION ALGORITHM

A lot of optimization methods have been developed for solving different types of optimization problems in recent years. There is no known single optimization method available for solving all optimization problems.

The modern optimization methods are very powerful and popular methods for solving complex engineering problems. These methods are particle swarm optimization algorithm, neural networks, genetic algorithms, artificial immune systems, and fuzzy optimization.

The particle Swarm concept originated as a simulation of simplified social systems. The original intent was to graphically simulate the graceful but unpredictable choreography of a bird flock. At some point in the evolution of the algorithm, it was realized that the conceptual model was, in fact, an optimizer. Through a process of trial and error, a number of parameters extraneous to optimization were eliminated from the algorithm, resulting in the simple original implementation.

The Particle Swarm Optimization algorithm is a novel population-based stochastic search algorithm and an alternative solution to the complex non-linear optimization problem. The PSO algorithm was first introduced by Dr. Kennedy and Dr. Eberhart in 1995 and its basic idea was originally inspired by simulation of the social behavior of animals such as bird flocking, fish schooling and so on. It is based on the natural process of group communication to share individual knowledge when a group of birds or insects search food or migrate and so forth in a searching space, although all birds or insects do not know where the best position is. But from the nature of the social behavior, if any member can find out a desirable path to go, the rest of the members will follow quickly [6] [7].

As compared with other optimization methods, it is faster, cheaper and more efficient. In addition, there are few parameters to adjust in PSO. That's why PSO is an ideal optimization problem solver in optimization problems. PSO is well suited to solve the non-linear, non-convex, continuous, discrete, integer variable type problems.

In ACC_PSO, each member of the population is called a particle and the population is called a swarm. Starting with a randomly initialized population and moving in randomly chosen directions, each particle goes through the searching space and remembers the best previous positions, velocity and accelerations of itself and its neighbors. Particles of a swarm communicate good position, velocity and acceleration to each other as well as dynamically adjust their own position, velocity and acceleration derived from the best position of all particles. The next step begins when all particles have been moved. Finally, all particles tend to fly towards better and better positions over the searching process until the swarm move to close to an optimum of the fitness function as shown in fig1.

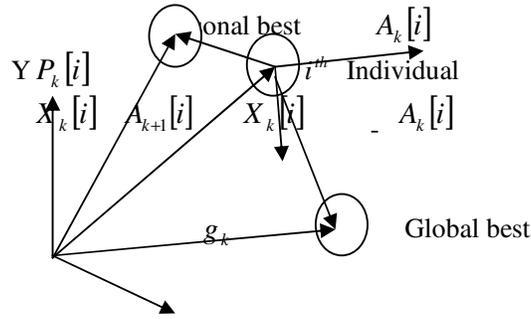

Figure1. Behavior of a individual in 2-dimensional search

ACCC_PSO has been used as a robust method to solve optimization problems in a wide variety of applications. In ACC_PSO for the optimization we have considered three parameters position, velocity and acceleration for each swarm particle, where as in PSO only two parameters position and velocity are considered for each particle. Here in this algorithm the swarms are the random sequence and rand positions are generated. From these positions, the velocity and acceleration are generated.

The following steps are followed to optimize the sequence.

1. Generate the individuals $x_o[i], i \in [0,2,....N-1]$ of initial generation (k=0) randomly.

2. For each particle, compute the position, velocity and acceleration and update $p_x[i], V_x[i], A_x[i]$ for all individuals.

3. Compute the level for each particle depending on the weigh of each particle. The new vector generated is considered as swarm particle.

4. Compute the best acceleration for each particle and the best acceleration for all particles and called as local best and global beat in the iterations.

5. Update the new acceleration and add it to the swarm particle and get the new particle.

$$x_{k+1}[i] = x_k[i] + A_k[i], \forall i$$
$$A_k[i] = K.\{A_{k-1}[i] + c_1.\Delta(g - x_k[i]) + c_2.\omega(p_k(i) - x_k[i])\}, \forall i \quad (7)$$

Where, $c_1, c_2 \text{ and } K \in R$ are weighting coefficients.

$$\Delta = diag[\alpha_1, \alpha_2, .........\alpha_d]$$
$$\omega = diag[\beta_1, \beta_2, .........\beta_d] \quad \text{Where} \quad \alpha_i \in [0,1], \beta_i \in [0,1] \quad \text{where i is an pseudorandom numbers.}$$

6. After updating all the particles convert the level vector to the weigh of each particle and compute Cost Function mentioned in equation (6). If the fitness function is satisfied the process ends other wise the whole process is repeated from step 3.

## 4. DESIGN RESULTS

Based on the proposed algorithm in section 3, the Discrete Frequency Coded Waveform (DFCW) set with the length of 128 and code sets of 3 are designed. In this paper the autocorrelation and cross correlation are normalized with respect to sequence length. The Optimized sequences of DFCW_FH, DFCW_LFM and Modified DFCW_LFM waveforms are generated using ACC_PSO. The simulation is carried out in Matlab. DFCW pulses have different starting frequencies with different combinations of BT, T.Δf and B/Δf. The Table1 shows the relation between the various parameters. From Table 2 to 4 show the ASPs and CPs for DFCW_FH, DFCW_LFM and Modified DFCW_LFM for the sequences generated by ACC_PSO. From the tabulated results it can be observed that the results ASPs and CPs of modified DFCW_LFM are better than that of DFCW_LFM and DFCW_FH. The Autocorrelation and cross correlation functions for the 3 waveforms of DFCW_LFM for the sequence are shown in Fig2 and Fig 3 respectively.

The effect of Doppler on the designed DFCW_LFM waveform is considered. The output of the matched filter of the designed DFCW_LFM is reduced by considering $f_d T = 0.031$ [2]. The Partial ambiguity function for one designed waveform is plotted in the Fig 4 and the complete ambiguity function for the designed waveform is plotted in Fig 5. Convergence of cost function for Genetic Algorithm and Particle Swarm Optimization algorithm and ACC_PSO is plotted in Fig 6. The convergence of ACC_PSO is faster. Fig 7 & 8 shows that as the number of sequence length increases the correlation properties decreases for both the algorithms. Fig 9 & 10 shows that as the number of antennas increases the correlation properties does not change much for both the algorithms. The correlation properties are better for ACC_PSO than GA as the number of sequence length and number of antennas increases.

Using ACC_PSO algorithm the comparison is made with different waveforms like DFCW_FH, DFCW_LFM and Modified DFCW_LFM in terms of autocorrelation sidelobe peak and cross correlation peak values (in dB) in Table 6. The results show an improvement in ASPs and CPs. It infers that sequences generated by ACC_PSO algorithm have good correlation properties with Modified DFCW_LFM.

Table 2 ASPs and CPs of the designed DFCW_FH waveform using ACC_PSO for N=128 L=3

|  | Code1 | Code2 | Code3 |
|---|---|---|---|
| Code1 | 0.076 | 0.0145 | 0.0141 |
| Code2 | 0.0145 | 0.087 | 0.0140 |
| Code3 | 0.0141 | 0.0140 | 0.053 |

Table 3 ASPs and CPs of the designed DFCW_LFM waveform using ACC_PSO for N=128, L=3

|  | Code1 | Code2 | Code3 |
|---|---|---|---|
| Code1 | 0.043 | 0.0101 | 0.009 |
| Code2 | 0.0101 | 0.057 | 0.0098 |
| Code3 | 0.009 | 0.0098 | 0.03 |

Table 4 ASPs and CPs of the designed Modified DFCW_LFM waveform using ACC_PSO for N=128, L=3

|  | Code1 | Code2 | Code3 |
|---|---|---|---|
| Code1 | 0.0280 | 0.0077 | 0.0087 |
| Code2 | 0.0077 | 0.0428 | 0.0088 |
| Code3 | 0.0087 | 0.0088 | 0.097 |

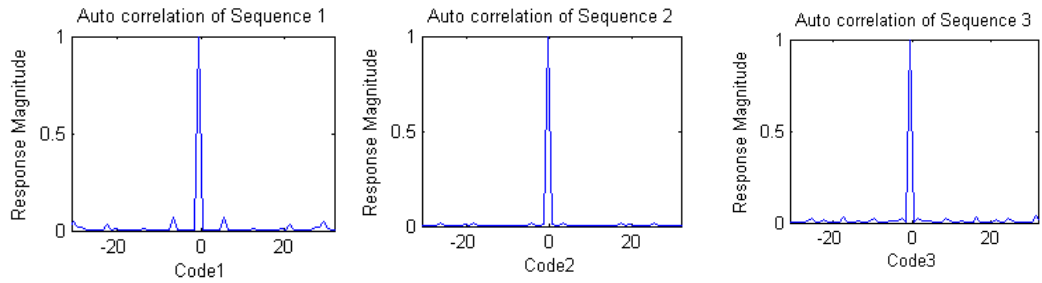

Figure 2. Auto Correlation functions of sequence length N=128 and Antenna=3.

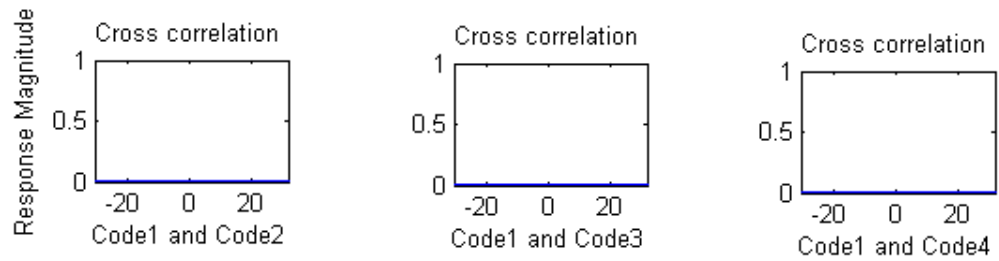

Figure 3. Cross Correlation functions of sequence length N=128 and Antenna=3.

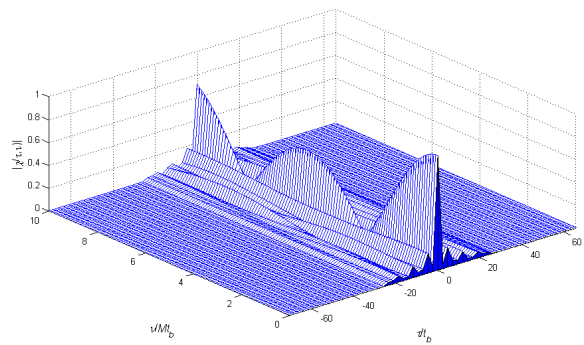

Figure 4. Partial Ambiguity function for sequence length N=32 at one receiver.

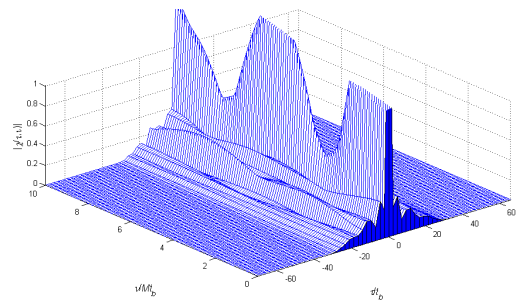

Figure 5. Total Ambiguity function for sequence length N=32 at one receiver

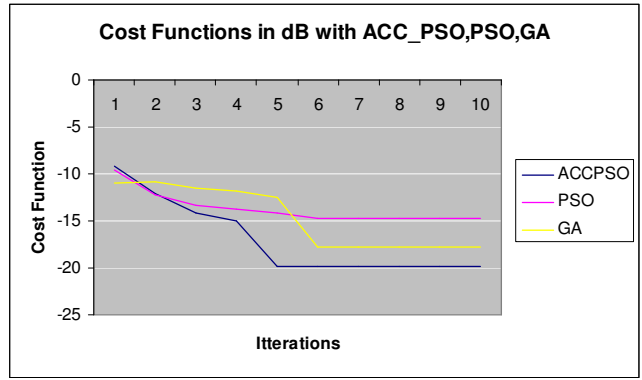

Figure 6.  Convergence of Cost Function in dB.

Table 6 ASPs and CPs for different waveforms in dB using ACC_PSO for N=128, L=3

| Algorithm | ASPs result in dB | CPs result in dB |
|---|---|---|
| ACC_PSO DFCW_FH | -26.02 | -37.7 |
| ACC_PSO DFCW_LFM | -30.4 | -40.9 |
| ACC_PSO Modified DFCW_LFM | -31.05 | -42.2 |

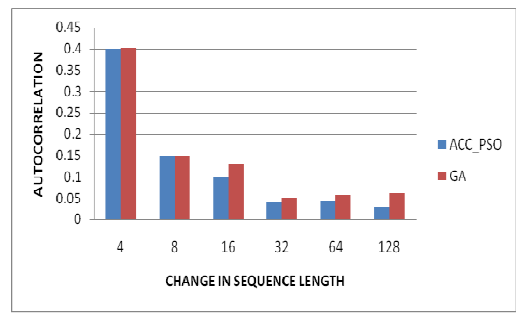

Figure 7.  Autocorrelation Vs sequence length for DFCW waveform..

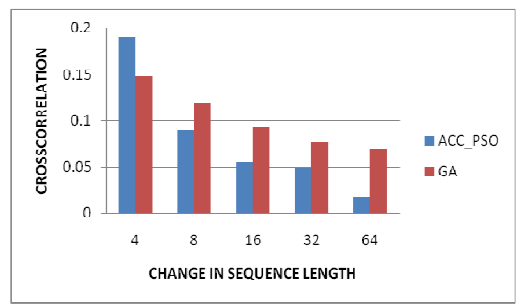

Figure 8.   Crosscorrelation Vs sequence length for DFCW waveform.

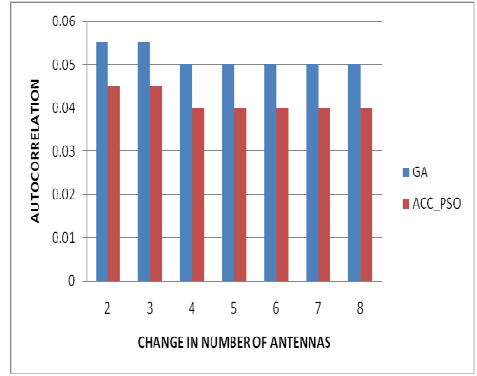

Figure 9.  Autocorrelation Vs antennas for DFCW waveform.

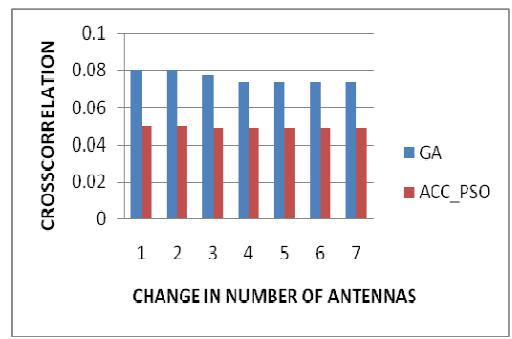

Figure 10.  Crosscorrelation Vs antennas for DFCW waveform.

## 5. CONCLUSIONS

In this paper, we have presented a new numerically optimized method of discrete frequency-coding waveform for orthogonal MIMO radar. This method is applicable to the case where the transmitted waveforms are orthogonal. The proposed method applies the Accelerated particle swarm optimization algorithm can get superior correlation properties. This approach is an alternative tool for the design of multiple orthogonal discrete frequency coding sequences with good correlation. The effectiveness of the proposed algorithm was demonstrated through the design results.

**Authors**

**Smt. B. Roja Reddy** received the B.E degree in 1998 from Gulbarga

University, Karnataka and M.E degree in 2004 from VTU, Karnataka. Presently working at R.V. college of Engineering with an experience of  9 years in the teaching field. Her research interest lies in various areas signal Processing. Currently précising her Ph.D in Radar Signal Processing.

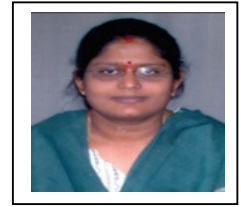

**Dr. M Uttara Kumari** received the B.E degree in 1989 from Nagarujna University, Hyderabad, Andra Pradesh and M.E degree in 1996 from Bangalore University,   Karnataka and  Ph.D degree in 2007 from Andhra University. Presently working at R.V. college of Engineering with an experience of  17 years in the teaching field. Her research interest lies in various areas of  radar systems, Space-time adaptive processing, speech processing and image processing**.**

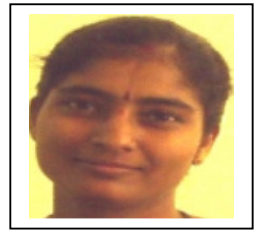